\documentclass{article}
\usepackage{spconf}
\usepackage{amsmath}
\usepackage{bm}
\usepackage{braket}
\usepackage{cite}
\usepackage{graphicx}

\title{Neural Sequence-to-Sequence Speech Synthesis Using a Hidden Semi-Markov Model
Based Structured Attention Mechanism}
\name{
\begin{tabular}{c}
  Yoshihiko Nankaku, Kenta Sumiya, Takenori Yoshimura, Shinji Takaki,\\
   Kei Hashimoto,  Keiichiro Oura and Keiichi Tokuda
\end{tabular}
}

\address{Nagoya Institute of Technology, Japan}

\begin{document}
\maketitle

\begin{abstract}
This paper proposes a novel Sequence-to-Sequence (Seq2Seq) model
integrating the structure of Hidden Semi-Markov Models (HSMMs) into
its attention mechanism.
In speech synthesis, it has been shown that methods based on Seq2Seq
models using deep neural networks can synthesize high quality speech
under the appropriate conditions. 
However, several essential problems still have remained, i.e., requiring
large amounts of training data due to an excessive degree for freedom
in alignment (mapping function between two sequences), and the
difficulty in handling duration due to the lack of explicit duration
modeling. 
The proposed method defines a generative model to realize the
simultaneous optimization of alignments and model parameters based on 
the Variational Auto-Encoder (VAE) framework, and provides monotonic
alignments and explicit duration modeling based on the structure of
HSMM. 
The proposed method can be regarded as an integration of Hidden Markov
Model (HMM) based speech synthesis and deep learning based speech
synthesis using Seq2Seq models, incorporating both the benefits.
Subjective evaluation experiments showed that the proposed method
obtained higher mean opinion scores than Tacotron~2 on relatively
small amount of training data. 
\end{abstract}

\begin{keywords}
Speech Synthesis, Deep Neural Networks, Attention Mechanism, Hidden 
Semi-Markov Models, Sequence-to-Sequence Models
\end{keywords}

\section{Introduction}
There has been much recent research on end-to-end text-to-speech
synthesis based on neural networks.  
In essence, text-to-speech synthesis is a sequence transform
generating a sequence of acoustic features from a sequence of
characters. Therefore, it is well-suited to using a
Sequence-to-Sequence (Seq2Seq) model with an attention mechanism that  
infers the relationship (alignment) between the sequences
\cite{taco2}.  
Although it has been shown that methods based on Seq2Seq models can
synthesize high quality speech under the appropriate conditions,
critical problems still have remained, i.e., requiring large amounts
of training data due to an excessive degree for freedom in alignment
obtained from the attention mechanism, and the difficulty in handling
duration because explicit alignment information cannot be obtained. 
To overcome these problems, some improved techniques have been proposed
recently, e.g., applying constraints to attention which tends to yield
monotonic alignment \cite{f-att,guided}, and incorporating explicit
alignment and/or duration to Seq2Seq models \cite{fast-sp, align-tts,
  vq-vae}.  
However, no method has yet been established that can perform overall
optimization of both alignment and model parameters while also
considering duration.  

An important fact to be noticed is that above mentioned problems were
appropriately solved in the conventional speech synthesis based on
hidden Markov models (HMM-based speech synthesis). 
HMM-based speech synthesis uses a hidden semi-Markov model (HSMM), which is
an HMM incorporating a state duration model, with state sequence as a
latent variable representing alignment, and performing simultaneous
optimization while also considering duration. 
In this paper, we propose a speech synthesis technique based on a 
Seq2Seq model with attention mechanism incorporating an HSMM
structure. The proposed method is composed of a generative model based
on a variational auto-encoder (VAE) \cite{vae}, in which the alignment 
between input and output sequences is represented as latent variables
as HSMM. The proposed method can also be interpreted as one of forms
of structured attention [8] using HSMM.
The alignment obtained by the proposed method is monotonic and
consistent over an entire sequence, therefore it is expected to be
able to build high quality systems using less training data than the
conventional Seq2Seq models. 
Moreover, since duration is handled explicitly in the proposed model,
it can be controlled in speech synthesis. 
The proposed method can be regarded as an integration of previous
HMM-based speech synthesis and recent deep learning based speech
synthesis using Seq2Seq models, incorporating both the benefits.

\section{Related Work}
\subsection{Seq2Seq models with an attention mechanism}
Sequence-to-Sequence models using an attention mechanism
\cite{seq2seq} can learn the time correspondence relationship between
two sequences of different lengths, and have been used in end-to-end
speech synthesis models such as Tacotron~2 \cite{taco2}. 
These models are composed of three main elements: an encoder, a
decoder, and an attention mechanism. 
The attention mechanism probabilistically selects an encoder hidden
state for each decoder time step,  
which enables it to obtain suitable latent representations for
sequences of various lengths. 
The context vector $\bm{c}_i$, obtained by the attention
mechanism for time $i$ is represented by a weighted sum of encoder
hidden states $\bm{h}_j$ as follows. 
\begin{equation}
  \bm{c}_i = \sum_{j=1}^{N} \alpha_{ij}\bm{h}_j \label{att-dist}
\end{equation}
where $\sum_j \alpha_{ij} = 1$, $N$ is the input sequence length, and
$\alpha_{ij}$ are probabilities representing the degree of attention on
the $j$th hidden state $\bm{h}_j$ in the encoder, 
computed from $\bm{h}_j$ and the $i-1$th state in the decoder. 
Attention corresponds to the alignment between linguistic and acoustic feature
sequences in speech synthesis, and can be obtained automatically
through training. 
However, the above mentioned attention mechanism has excessive degree
of freedom in matching function between two sequence and requires
suitable constraints to keep monotonic alignment in speech synthesis.
Moreover, estimating matching function does not take into account of
the duration of units in linguistic feature sequences, e.g., phones.

\subsection{DNN-HSMM} 
A hidden semi-Markov models (HSMM) is a HMM incorporating a duration
model, and its likelihood function is given as follows.
\begin{eqnarray}
\lefteqn{
  p(\bm{o} \mid \bm{l}, \bm{\lambda}_{\text{HSMM}})  = \sum_{\bm{z}}
  p(\bm{o}, \bm{z} \mid  \bm{l})
}\\
&=& \sum_{\bm{z}} \left\{  \prod_{t=1}^T p(\bm{o}_t \mid z_t,  \bm{l}) \prod_{k=1}^K p(d_k \mid  \bm{l})   \right\} 
\end{eqnarray}
where $\bm{o}=(\bm{o}_1, \bm{o}_2, \ldots, \bm{o}_T)$, and $\bm{l}=(\bm{l}_1,\bm{l}_2, \ldots,\bm{l}_K)$
represent acoustic and linguistic feature values respectively, and $\bm{z}$ represents
the state sequence for alignment. State duration, $d_k$ represents the
duration of each state, $k$ in $\bm{z}$.
\begin{eqnarray}
  \bm{z} & =&\left( z_{1},  z_{2}, \ldots, z_{T}  \right) \\
          & = &(\underbrace{S_1,\ldots,S_1}_{\times d_1},
                \underbrace{S_2, \ldots, S_2}_{\times d_2}, \ldots,
                \underbrace{S_K, \ldots, S_K}_{\times d_K})
 \end{eqnarray}
Usually the parameters of HSMM are optimized by the
expectation-maximization (EM) algorithm for the maximum likelihood
estimates. The algorithm iteratively updates the the posterior
probability distribution $p(\bm{z} \mid \bm{o},\bm{l})$ and model
parameters $\bm{\lambda}_{\text{HSMM}}$, and the posterior probability 
can be efficiently computed using a generalized forward-backward
algorithm \cite{gfb}. 
A DNN-HSMM \cite{dnn-hsmm} has been proposed as a statistical model 
for incorporating the qualities of an HSMM, i.e., handling duration
appropriately, into a neural network. 
In the DNN-HSMM, the neural network takes linguistic features as
input and generates HSMM model parameters. Taking both output and
duration probabilities as Gaussian distributions, this yields:
\begin{eqnarray}
p(\bm{o}_t \mid z_t=k,  \bm{l}) &=& {\cal N}(\bm{o}_t \mid \bm{\mu}_{k}, \bm{\sigma}^{2}_{k})\\
p(d_k \mid  \bm{l}) &=&  {\cal N}(d_k \mid {\xi}_{k}, {\eta}_{k}^2) \\
  \left\{ \bm{\mu}_{k}, \bm{\sigma}^{2}_{k}, {\xi}_{k}, {\eta}_{k}^2
  \right\} &=& \text{DNN}(\bm{l}_k)
\end{eqnarray}
This achieves a more flexible mapping from linguistic features to
model parameters than the decision-tree-based clustering used in
HMM-based speech synthesis, while duration and alignment can be
appropriately handled within the framework of statistical generative 
models. 
However, due to the limitation that final acoustic features $\bm{o}$
are generated by the HSMM, quality is somewhat inferior compared with
recent Seq2Seq models. 

\section{Proposed Method}
\subsection{HSMM based structured attention}

\begin{figure}[t]
%  \begin{minipage}[b]{1.0\linewidth}
%    \centerline{\includegraphics[width=8.5cm]{./figure/hsmm-attention.pdf}}
  \includegraphics[width=8.5cm]{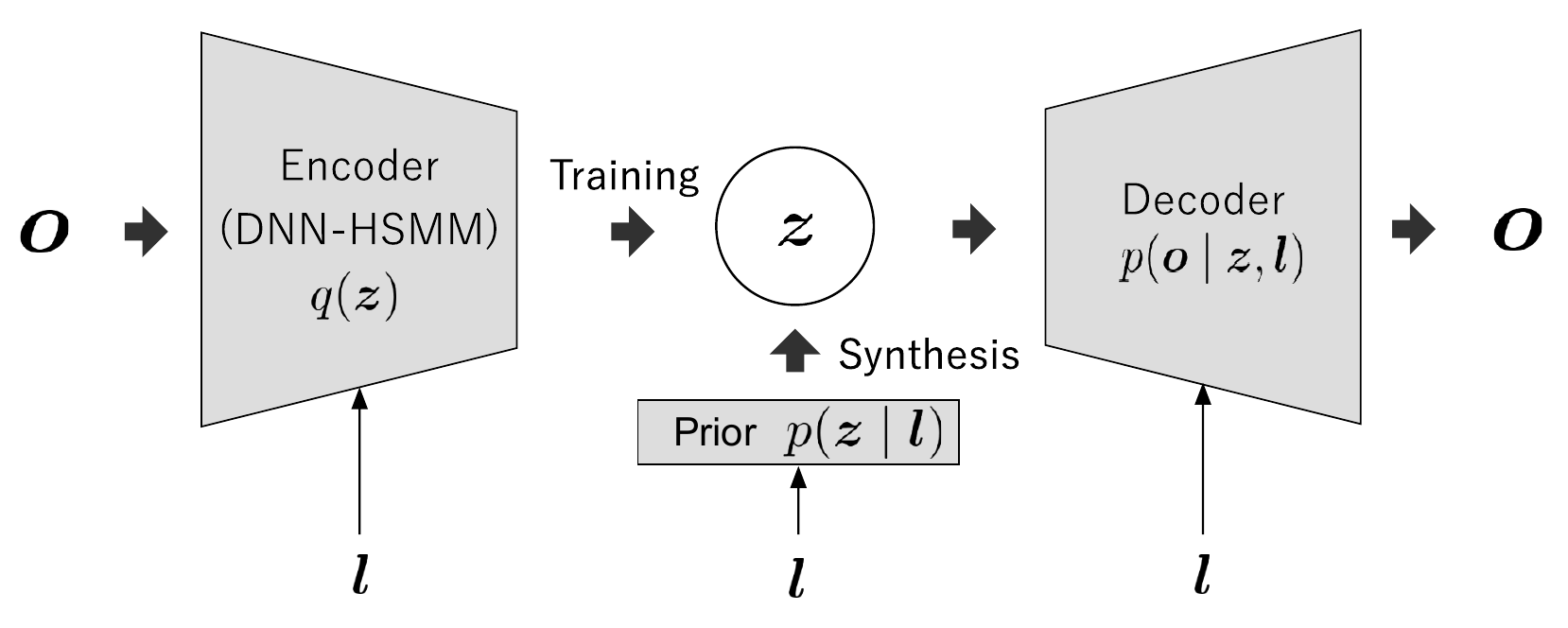}
    \caption{VAE based structured attention}
    \label{fig:HSMM-attention}
%  \end{minipage}
\end{figure}

\begin{figure*}[t]
  \begin{center}
    \includegraphics[width=16cm]{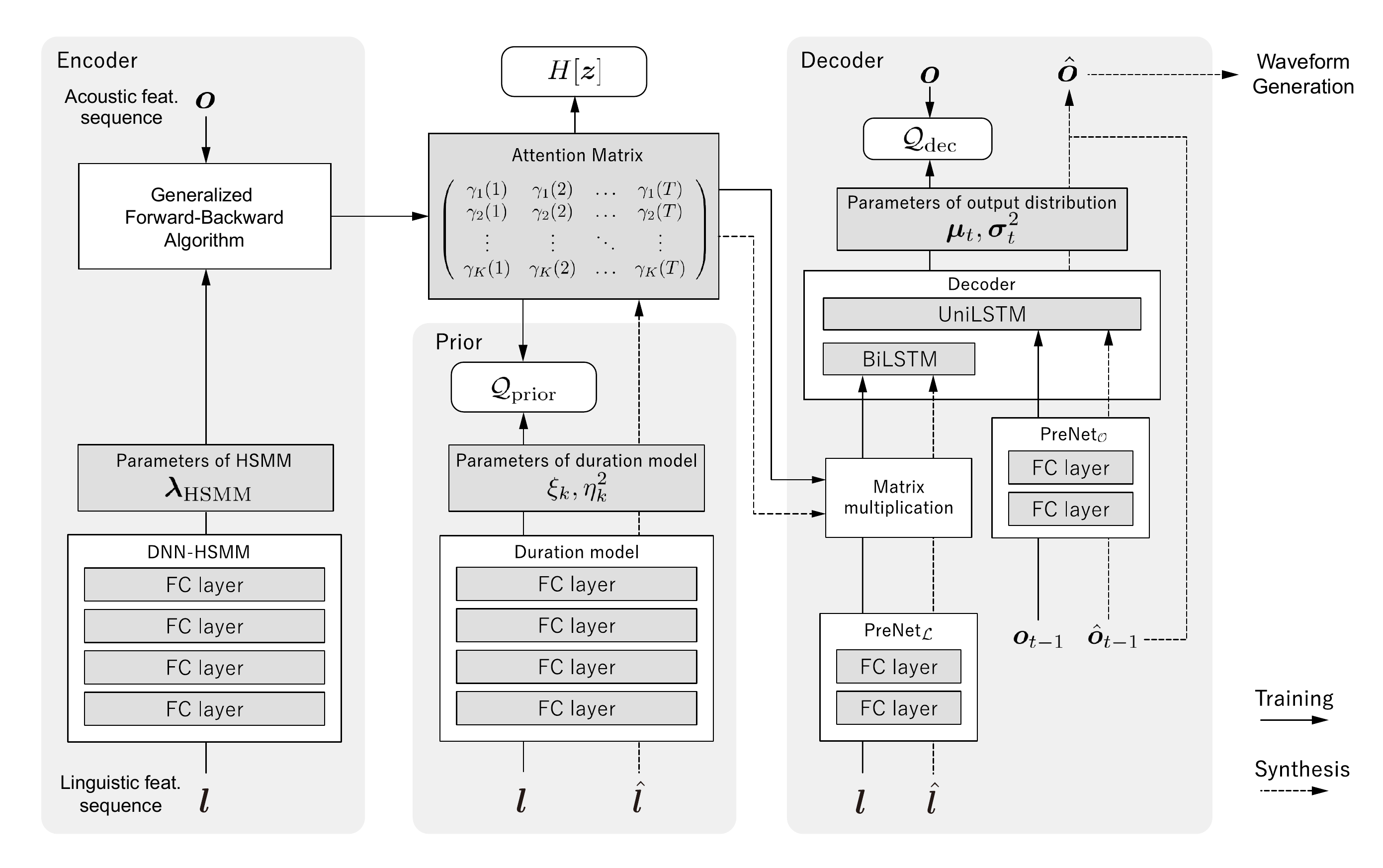}
    \caption{Proposed model structure}
    \label{fig:proposed}
  \end{center}
\vspace{-5mm}
\end{figure*}

This paper proposes a novel Seq2Seq model based on an attention
mechanism integrating the structure of HSMM.
The method defines a generative model based on a variational
auto-encoder (VAE) \cite{vae} framework, in which the alignment
between acoustic feature and linguistic feature sequences is
represented by the discrete latent variable sequence corresponding to
HSMM states.  
An overview of the method is shown in Figure \ref{fig:HSMM-attention}. 
We first define an evidence lower bound (ELBO) based on the likelihood
function $p(\bm{o} \mid \bm{l})$; 
\begin{eqnarray}
\lefteqn{  \log p(\bm{o} \mid \bm{l}) 
  = \log \sum_{\bm{z}}  p(\bm{o}, \bm{z} \mid \bm{l})} \\
&\geq& \sum_{\bm{z}}  q(\bm{z}) 
       \log \frac{p(\bm{o}, \bm{z} \mid \bm{l})}{q(\bm{z})}   \label{eq:jneq2} \\ 
&=&    \sum_{\bm{z}}  q(\bm{z}) \log p(\bm{o} \mid \bm{z},\bm{l}) 
    +  \sum_{\bm{z}}  q(\bm{z}) \log p(\bm{z} \mid \bm{l})\nonumber \\
&& -  \sum_{\bm{z}}  q(\bm{z}) \log q(\bm{z}) \label{eq:reconst} \\
&=&  \mathcal{L}_{\text{ELBO}}
\label{eq:ELBO}
\end{eqnarray}
where $q(\bm{z})$ and $p(\bm{o}\mid\bm{z},\bm{l})$ are the encoder
and decoder respectively, composed of separate neural
networks. Usually, prior distribution $p(\bm{z} \mid \bm{l})$ is set
to a particular distribution beforehand, e.g., a normal distribution.
In the proposed method, we assume it consists of a neural network that
will be trained.  

The VAE optimizes the neural network by maximizing the evidence lower
bound $\mathcal{L}_{\text{ELBO}}$. 
In the proposed model, latent variable $\bm{z}$ represents a discrete
state sequence similarly to an HSMM, and the approximate posterior
distribution $q(\bm{z})$ represented by the encoder is assumed to be
composed of a DNN-HSMM: 
\begin{eqnarray}
  q(\bm{z})                & = & p(\bm{z} \mid \bm{o}, \bm{\lambda}_{\text{HSMM}}) \\
  \bm{\lambda}_\text{HSMM} & = & \text{DNN-HSMM}(\bm{l})
\end{eqnarray}
In other words, the proposed method uses the HSMM as an estimator
for state posterior probability $q(\bm{z})$  while the conventional
DNN-HSMM uses it as a generator of outputs acoustic features $\bm{o}$. 
Normally, to calculate the ELBO (equation (\ref{eq:ELBO})), sampling
the latent variables from the posterior distribution is used in VAEs.
However, since $\bm{z}$ is a discrete variable sequence in the
proposed method, the following expectation is used to propagates
encoder information to the decoder
\begin{equation}
  \gamma_k(t) = p(z_t=k \mid \bm{o}, \bm{l}) = \sum_{\bm{z}}q(\bm{z})\delta(z_t,k)
\end{equation}
It can be clearly seen that $\gamma_k(t)$ is the attention itself,
that is, $\gamma_k(t)$ represents the degree of attention on
linguistic feature $\bm{l}_k$ when generating the frame for time $t$. 
Even though the calculation of $\gamma_k(t)$ requires counting all
possible state sequences, it can be efficiently calculated using a
generalized forward-backward algorithm as in DNN-HSMM.  
This enables the proposed method to estimate attention based on
consistent alignment over the entire sequences representing monotonic
alignment.  

From the definition of ELBO, the decoder can be defined as a neural
network with any structure whose inputs are $\bm{z}$ and $\bm{l}$, and
output is $\bm{o}$. However, as mentioned above, assuming expectation 
$\bm{\gamma}=\left\{\gamma_k(t) \mid k=1,\ldots,K,t=1,\dots,T\right\}$ instead of
$\bm{z}$ is passed from the encoder,  
it means that the decoder has the following approximation.
\begin{eqnarray}
  \sum_{\bm{z}}  q(\bm{z}) \log p(\bm{o} \mid \bm{z},\bm{l}) 
%& = \Braket{ \log p(\bm{o} \mid \bm{z},\bm{l})}_{q(\bm{z})} \\
& \approx & 
\log p(\bm{o} \mid  \bm{\gamma}, \bm{l} )
\end{eqnarray}
Moreover, by applying the structure similar to a conventional attention
mechanism, the exact same form as the decoder in Seq2Seq models:
\begin{eqnarray}
\log p(\bm{o} \mid  \bm{\gamma}, \bm{l} ) & = & \log p(\bm{o} \mid  \Braket{\bm{l}} ) \\
  \Braket{\bm{l}} & = & (\Braket{\bm{l}_1},\Braket{\bm{l}_2},\ldots, \Braket{\bm{l}_T})  \\
  \Braket{\bm{l}_t} & = & \sum_{k=1}^{K} {\gamma_k{(t)}}f(\bm{l}_k) \label{hsmm-att}
\end{eqnarray}
where $f(\cdot)$ is the decoder PreNet (corresponding to the encoder in the
Seq2Seq model), and $\Braket{\bm{l}_t}$ is the context vector. 
To perform the back-propagation to the encoder parameters, 
although the original VAEs use the re-parameterization trick \cite{vae},
the proposed method can directly apply the back-propagation to the
encoder parameters through expectation $\bm{\gamma}$, because $\bm{\gamma}$ are
composed of encoder parameters (HSMM).

The detailed structure of the proposed model is shown in Figure \ref{fig:proposed}. 
In the figure,
$\mathcal{Q}_{\text{dec}}$, $\mathcal{Q}_{\text{prior}}$ and
$H[\bm{z}]$ correspond to the terms in Equation (9), and are computed
as follows. 
\begin{eqnarray}
  \mathcal{Q}_{\text{dec}}  &=& \log p(\bm{o} \mid  \Braket{\bm{l}} ) \\
  \mathcal{Q}_{\text{prior}} &=& \sum_{t=1}^T\sum_{d=1}^{D}\sum_{k=1}^K \gamma_k^{(d)}(t)\log p(d \mid \bm{l}_k) \\
  H[\bm{z}]                &=& -\sum_{t=1}^T\sum_{k=1}^K  {\gamma_k}(t)\log{\gamma_k}(t)
\end{eqnarray}
where $\gamma_k^{(d)}(t)$ is the probability that state $k$ is
continuously selected in the time interval from $t-d$ to $t$, 
and can be computed similarly to $\gamma_k(t)$ by using the
generalized forward-backward algorithm. 
It is assumed that decoder $p(\bm{o} \mid \Braket{\bm{l}})$ 
and prior distribution $p(d \mid \bm{l}_k)$ are Gaussian
distributions whose parameters are generated from the corresponding
neural networks,  $\text{Decoder}(\cdot)$ and  $\text{Prior}(\cdot)$,
respectively. 
\begin{eqnarray}
  p(\bm{o} \mid \Braket{\bm{l}})      & = & \prod_{t=1}^T{\cal N}(\bm{o}_t \mid \bm{\mu}_{t}, \bm{\sigma}^{2}_{t}) \\
  \left\{ \bm{\mu}_{t}, \bm{\sigma}^{2}_{t} \right\}  & = & \text{Decoder}( \Braket{\bm{l}_t}, h(\bm{o}_{t-1}))   \\
  p(d_k \mid  \bm{l}_k)               & = & {\cal N}(d_k \mid {\xi}_{k}, {\eta}_{k}^2) \\
  \left\{ {\xi}_{k}, {\eta}_{k}^2 \right\}            & = & \text{Prior}(\bm{l}_k)
\end{eqnarray}
For the decoder, even though any type of decoder can be used,
e.g. LSTM or a recent feed-forward Transformer \cite{fast-sp, align-tts}, 
this paper used an autoregressive structure for comparison with
Tactron~2. 
In the equation, $h(\cdot)=\text{PreNet}_{\mathcal{O}}(\cdot)$
represents the auto-regression PreNet. The prior distribution is
configured as a neural network separate from the encoder, but it
shares parameters with the encoder, and the duration distribution can
also be used as the prior distribution in the HSMM.

During the synthesis stage, the likely state sequence is computed from
prior distribution and encoder by:
\begin{equation}
  \hat{\bm{z}} = \arg \max_{\bm{z}} \prod_{k=1}^K {\cal N}(d_k \mid {\xi}_{k}, {\eta}_{k}^2) \\
\end{equation}
The decoder generates acoustic features, driven using attention
composed of this $\hat{\bm{z}}$.

\section{Evaluation}
\subsection{Experimental Conditions}
To show the effectiveness of the proposed method, experiments were
conducted using speech data from a single male speaker. 
The speech database consists of 503 Japanese sentences; 450 sentences
were used as training data, and remaining 53 sentences were used as
test data. The sampling rate of the speech data was 48 kHz. 
Three methods in addition to the proposed method (\textbf{PROPOSED}) were 
compared; DNN-HSMM (\textbf{DNN-HSMM}), Tacotron~2 (\textbf{TACO}), and
a frame-unit acoustic model (\textbf{FRAME}).  
The target acoustic features to be modeled included a 50-dimensional
mel-cepstrum coefficient vector extracted by STRAIGHT analysis
\cite{straight}, the log-fundamental frequency, V/UV, and a
25-dimensional aperiodicity features.  
For \textbf{DNN-HSMM} and \textbf{PROPOSED}, acoustic features with
their dynamic features were assumed to be generated from the HSMM. 
The HSMMs had five states, structured left-to-right with no
skips. 
As an input feature vector, a 386-dimensional phoneme-unit linguistic
feature was used for \textbf{TACO}, and a 388-dimensional state-unit
linguistic feature with a phoneme internal state index was used for
\textbf{DNN-HSMM} and \textbf{PROPOSED}. 
For \textbf{FRAME}, we used a 394-dimensional frame-unit linguistic feature
consisting of the state-unit linguistic feature, using \textbf{DNN-HSMM} for
alignment and adding items including a position number within the
state segment and duration context. During synthesis in
\textbf{FRAME}, the duration estimated by \textbf{DNN-HSMM} was used.
The model structure for \textbf{DNN-HSMM} is similar to the encoder
part of the proposed method (Figure \ref{fig:proposed}), and trained
based on the maximizing likelihood criterion.  
The model structure of \textbf{TACO} was adopted from the reference
\cite{taco2}, replacing the embedding layer with a linear transform
and adding guided attention loss ($g=0.2$) \cite{guided} for training. 
The reduction factor of 3 was used, and only monophone labels were
enabled in the initial stage of training. 
The decoder in the \textbf{FRAME} was the same as for the proposed
method (Figure \ref{fig:proposed}), 
and training was conducted with the frame-unit linguistic feature as
input to the $\text{PreNet}_{\mathcal{L}}(\cdot)$. From preliminary
test results, the prior distribution and encoder duration model were
shared in \textbf{PROPOSED}.  
In the training procedure for  \textbf{PROPOSED}, initial training of
the decoder was performed with setting the trained \textbf{DNN-HSMM}
to the encoder and then an overall optimization was
conducted. Adam\cite{adam} was used as the optimization algorithm for
all training in all methods. 

\subsection{Subjective Evaluation}

\begin{figure}[t]
  \begin{center}
    \includegraphics[width=8cm]{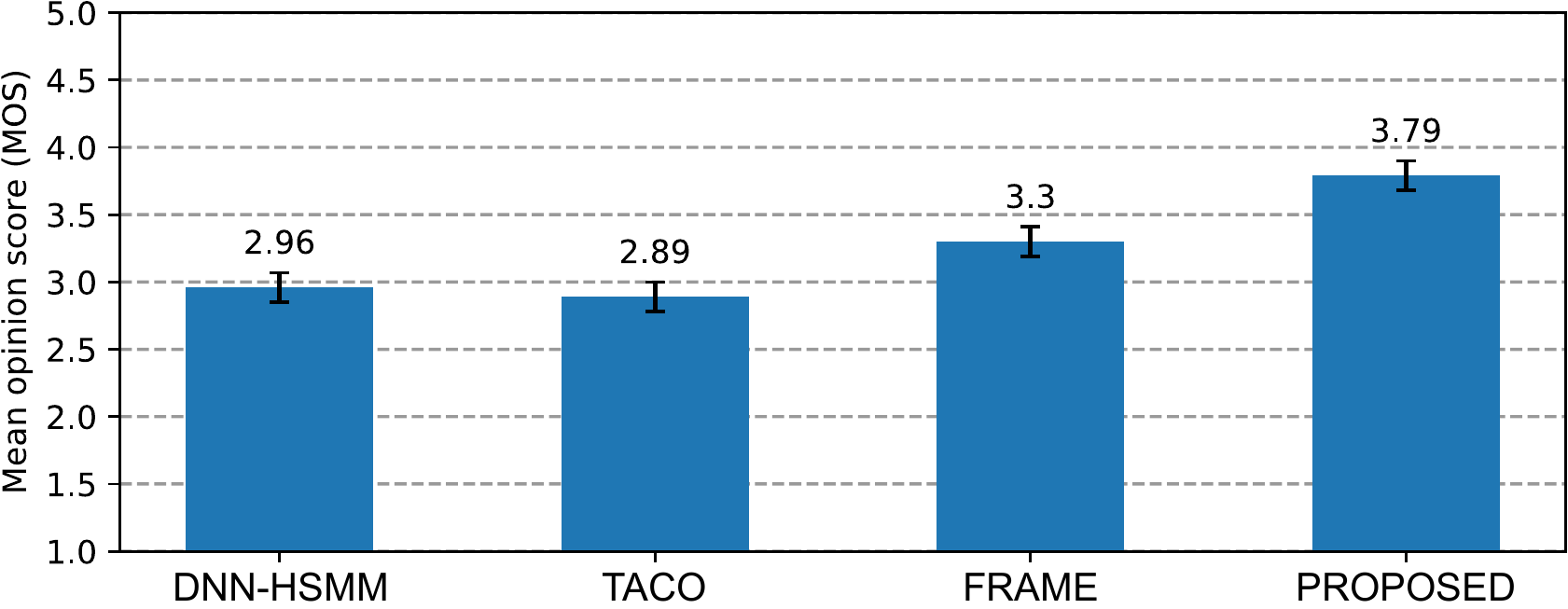}
  \end{center}
  \vspace{-10pt}
  \vspace{-3pt}
  \caption{Subjective evaluation results}
  \label{fig:subject}
\end{figure}

The naturalness of synthesized speech was evaluated using Mean Opinion
Score (MOS) tests. For each of ten subjects, 15 sentences from the test
data were selected at random and evaluated in 5-scale scores. 
The results of subjective evaluation are shown in Figure 3. 
The figure shows that the \textbf{PROPOSED} received the
highest score for naturalness, confirming that the method is
effective. In particular, \textbf{PROPOSED} obtained a higher score
than \textbf{FRAME} using the same decoder structure.
This suggests that the simultaneous optimization of alignment and
model parameters was effective in the proposed method. 
Comparing \textbf{PROPOSED} and \textbf{DNN-HSMM}, 
a clear improvement of \textbf{PROPOSED} suggests that 
the flexible autoregressive decoder structure contributed
significantly to improving naturalness. 
Because of the relatively small amount of training data,
\textbf{TACO} obtained poor alignment accuracy due to its over
degree of freedom in attention, 
and the proposed method was able to achieve suitable alignment due to
the suitable structured attention.

\section{Conclusion}
This paper proposed a method for Seq2Seq modeling using structured
attention based on HSMM. The proposed method is formulated based on a
VAE framework, and can be regarded as an integration form of the
conventional HMM speech synthesis and a neural network based Seq2Seq
model. Experimental results showed improvement in naturalness of
synthesized speech due to the simultaneous optimization of alignment
and model parameters based on VAE framework. 
The robust alignment estimation based on the appropriate structure
constraint using HSMM also contributes to the quality of synthesized
speech especially on the small amount of training data.
Future work includes evaluation using larger speech database and
application to fully end-to-end speech synthesis and to multiple 
speaker models.

\end{document}